\documentclass[aps,prx,twocolumn,final,letterpaper, superscriptaddress]{revtex4}

\usepackage{appendix}
\usepackage{graphicx}   
\usepackage{import}                         
\usepackage{epstopdf}
\usepackage{amsmath} 
\usepackage{bm}
\usepackage{amssymb}
\usepackage{quotes}
\usepackage{transparent}
\usepackage{dcolumn}
\usepackage{multirow}
\usepackage{cancel} 
\usepackage{mdframed}
\usepackage{color}
\usepackage{bm}
\usepackage{dsfont}
\usepackage{slashed}

\usepackage{braket}

\newcommand{\appropto}{\mathrel{\vcenter{
  \offinterlineskip\halign{\hfil$##$\cr
    \propto\cr\noalign{\kern.2pt}\sim\cr\noalign{\kern-2.5pt}}}}}

\begin{document}
\rmfamily

\newcommand{\MITrle}{Research Laboratory of Electronics, Massachusetts Institute of Technology, Cambridge, MA 02139, USA}
\newcommand{\MITphysics}{Department of Physics, Massachusetts Institute of Technology, Cambridge, MA 02139, USA}
\newcommand{\harvard}{Department of Physics, Harvard, Cambridge, MA 02138, USA}
\newcommand{\technion}{Solid State Institute, Technion, Haifa, Israel}

\title{Entangling extreme ultraviolet photons through strong field pair generation}

\author{Jamison Sloan}
\email{jamison@mit.edu}
\affiliation{\MITrle}

\author{Alexey Gorlach}
\affiliation{\technion}

\author{Matan Even Tzur}
\affiliation{\technion}

\author{Nicholas Rivera}
\affiliation{\harvard}
\affiliation{\MITphysics}

\author{Oren Cohen}
\affiliation{\technion}

\author{Ido Kaminer}
\affiliation{\technion}

\author{Marin Solja\v{c}i\'{c}}
\affiliation{\MITrle}
\affiliation{\MITphysics}

\noindent	

\begin{abstract}
Entangled photon pairs are a vital resource for quantum information, computation, and metrology. Although these states are routinely generated at optical frequencies, sources of quantum of light are notably lacking at extreme ultraviolet (XUV) and soft X-ray frequencies. Here, we show that strongly driven systems used for high harmonic generation (HHG) can become versatile sources of entangled photon pairs at these high frequencies. We present a general theory of photon pair emission from non-perturbatively driven systems, which we refer to as "strong field pair generation" (SFPG). We show that strongly driven noble gases can generate thousands of entangled pairs per shot over a large XUV bandwidth. The emitted pairs have distinctive properties in angle and frequency, which can be exploited to discriminate them from the background HHG signal. 
We connect SFPG theory to the three-step-model of HHG, showing that this pair emission originates from the impact of high frequency vacuum fluctuations on electron recombination. 
The light produced by SFPG exhibits attosecond Hong-Ou-Mandel correlations, and can be leveraged as a source of heralded single photon attosecond pulses. Our findings aid ongoing efforts to propel quantum optics into the XUV and beyond.

\end{abstract}

\maketitle

\section{Introduction}


One of the profound surprises of the twentieth century was that particles can become entangled with one another, leading to seemingly non-local correlations that evade description by classical physics \cite{bell1964einstein, clauser1969proposed, aspect1982experimental}. A major area of impact of these ideas is in quantum optics, which was transformed by the measurement of entangled photon pairs generated through nonlinear optical processes \cite{harris1967observation, kwiat1995new}. Since then, decades of work have harnessed entangled pairs to interrogate the fundamental quantum nature of light \cite{hong1987measurement}, and to move the needle on quantum metrology, imaging, and information \cite{jennewein2000quantum}. Nowadays, optical and infrared (IR) pair generation is routine, and these sources can even be purchased, off the shelf. 

There are, however, massive ranges of the spectrum where quantum optics is much less developed. An example of this is at high frequencies (UV -- X-ray and beyond), where quantum sources are scarce. Exceptions on the UV end include resonant semiconductor sources \cite{edamatsu2004generation}, and non-degenerate four wave mixing sources which have recently entangled UV and IR frequencies \cite{lopez2023tunable}. At much higher energies, facility scale hard X-ray sources enable parametric down conversion into X-ray pairs \cite{freund1969parametric, shwartz2011polarization, strizhevsky2021efficient}, and highly non-degenerate X-ray -- UV/optical pairs \cite{schori2017x, sofer2019observation}. More recent theoretical proposals have also investigated spontaneous pair emission from ions \cite{wang2022attosecond}, and X-ray pair generation in free electron lasers \cite{zhang2023entangled}. 
Despite this progress toward high frequency sources, there remains a great need for new mechanisms of pair generation from compact sources, especially at frequencies ranging from the extreme ultraviolet (XUV) through the so-called ``water window,'' which is critical to biological imaging.

A leading technique for creating coherent XUV radiation is high harmonic generation (HHG), in which a strong IR field incident on a sample induces the conversion of many pump photons into high harmonic frequencies. HHG has been explored across many platforms (gases \cite{mcpherson1987studies, ferray1988multiple, krause1992high}, solids \cite{ghimire2019high}, liquids \cite{luu2018extreme}, plasmas \cite{gaudiosi2006high}), producing light ranging from the XUV to soft X-ray regimes \cite{popmintchev2015ultraviolet}. HHG also makes possible attosecond pulse generation \cite{paul2001observation, sansone2006isolated}, which fuels a vast array of spectroscopy, photoionization, and interferometry experiments that probe the dynamics of electrons on their natural timescale \cite{villeneuve2018attosecond, corkum2007attosecond}. 
The majority of experiments are well-described by considering the role of light in HHG from a classical perspective. However, some selected works --- starting decades ago \cite{gauthey1997high, lagattuta1993radiation}, and followed by a more recent resurgence \cite{gonoskov2016quantum, bogatskaya2016spontaneous, bogatskaya2017spontaneous, gombkotHo2020high, gombkotHo2021quantum, gorlach2020quantum} --- have investigated quantum optical aspects of high harmonic generation. A few pioneering experiments have already begun to report quantum features in HHG \cite{tsatrafyllis2017high, lewenstein2021generation, harrison2023generation}, and yet other theoretical proposals \cite{pizzi2023light, tzur2023upconverting, gorlach2023high} may soon become possible. 
Even though attosecond quantum optics has been identified as a critical new domain \cite{ko2023quantum, lewenstein2022attosecond}, HHG has not been considered as an entangled pair source.

\begin{figure*}[t]
    \centering
    \includegraphics{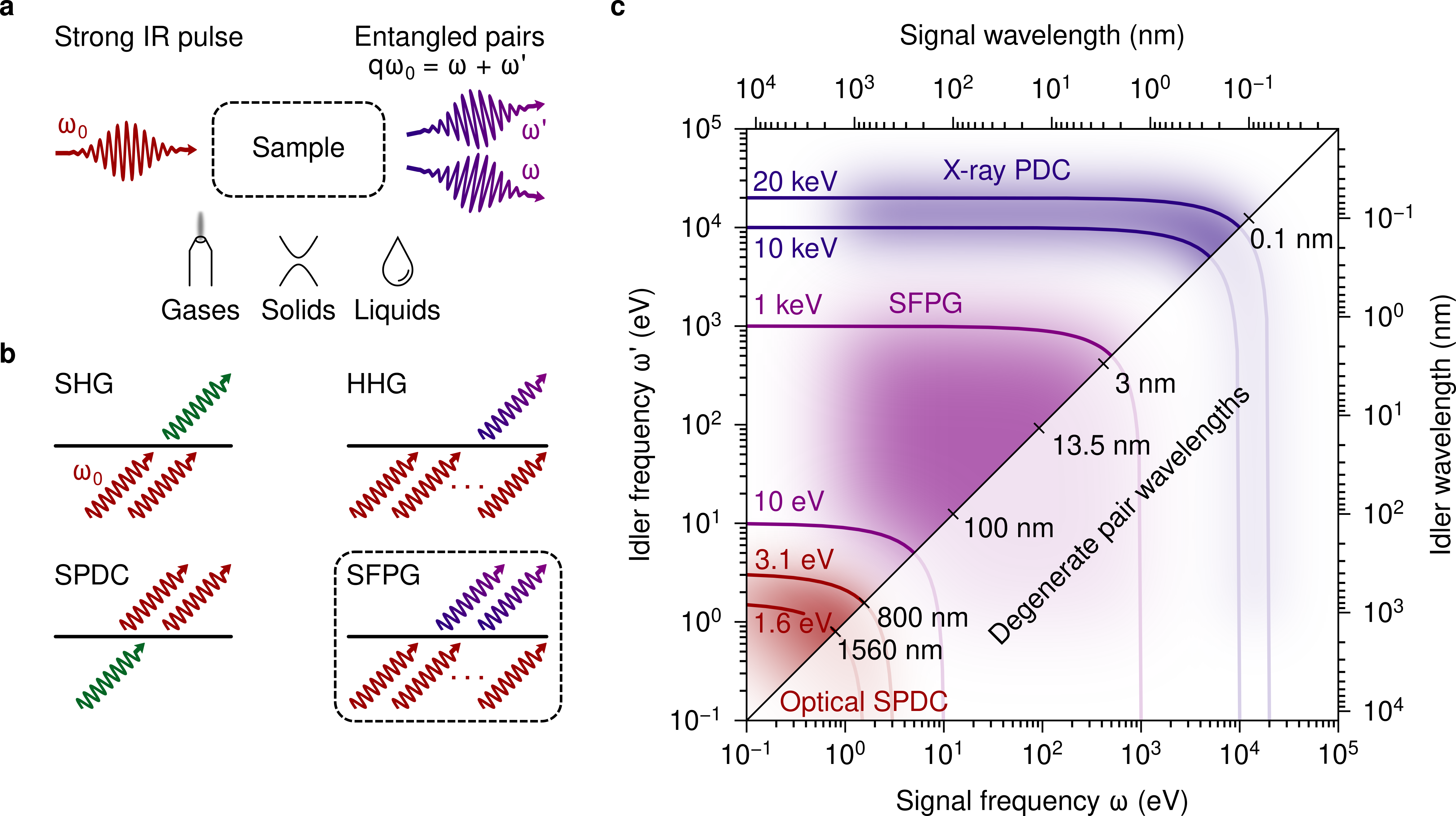}
    \caption{\textbf{Concept of strong field pair generation (SFPG).} \textbf{(a)} A strong infrared laser pulse of frequency $\omega_0$ is incident on a sample. When SFPG takes place, entangled photon pairs of frequencies $\omega$ and $\omega'$ are produced at angles away from the incident axis. \textbf{(b)} Feynman diagrams depict the quantum optical nature of various harmonic generation processes. The top row shows harmonic generation processes which result in the emission of a single photon: second harmonic generation (SHG) results from the conversion of two pump photons at frequency $\omega_0$ into a signal photon at $2\omega_0$; high harmonic generation (HHG) results from the conversion of $q$ pump photons into a high frequency signal photon at $q\omega_0$. The bottom row shows processes which generate entangled pairs: spontaneous parametric down conversion (SPDC) results from the conversion of one pump photon at $\omega_0$ into an entangled pair of photons $\omega$, $\omega'$ which satisfy $\omega + \omega' = \omega_0$; SFPG results from the conversion of $q$ pump photons into an entangled pair so that $\omega + \omega' = q\omega_0$. \textbf{(c)} State of current sources of degenerate and non-degenerate entangled photon pairs across a wide range of the electromagnetic spectrum. SFPG can produce highly tunable degenerate and non-degenerate pairs over a large range of the UV to soft X-ray where no other sources currently exist. 
    }
    \label{fig:schematic}
\end{figure*}



Here, we introduce the concept of ``strong field pair generation'' (SFPG), which enables broadly tunable sources of high frequency entangled pairs. 
In this non-perturbative quantum electrodynamical process, many low frequency pump photons in a high intensity field incident on a piece of material are converted into an entangled pair of photons at frequencies much higher than that of the pump (Figs.\ref{fig:schematic}a-b). 
We show that SFPG can yield degenerate XUV/X-ray pairs, or highly non-degenerate pairs which have an XUV signal but UV or optical/IR idler. Our results are based on a quantum optical theory of SFPG which predicts the spectral and angular correlations of entangled high frequency pairs, as well as the efficiency with which they can be generated. We interpret SFPG through the ``three-step-model'' of HHG, and show that the two-photon nature of SFPG leads to primary and secondary ``cutoff'' laws. The non-perturbative multi-harmonic nature of SFPG also leads to highly entangled biphoton states with attosecond correlation times, enabling heralded attosecond single photon sources. We show that SFPG should be observable within current HHG platforms, thus providing a new route to bring quantum optics to the attosecond domain.

Such XUV and soft X-ray entangled pairs would unearth many opportunities, both fundamental and applied. For example, such entangled pairs could lead to quantum-enhanced sensing modalities, such as two-photon spectroscopy or ghost imaging \cite{kutas2022quantum}. As another example, pair sources are known to exhibit squeezing of fluctuations below the famed ``shot noise limit,'' which could improve sensitivity in attosecond science. This is especially relevant in the context of recent experimental advances which achieve zeptosecond time resolution \cite{harrison2023generation}, and have potential to reach the yoctosecond scale, where it is anticipated that electromagnetic vacuum fluctuations influence electron dynamics \cite{even2023photon}. Moreover, sources of quantum XUV or soft X-ray radiation could probe high energy electronic or even low frequency nuclear transitions \cite{beeks2021thorium}, allowing a quantum optical interface to new degrees of freedom.

\section{Results}


\begin{figure*}
    \centering
    \includegraphics{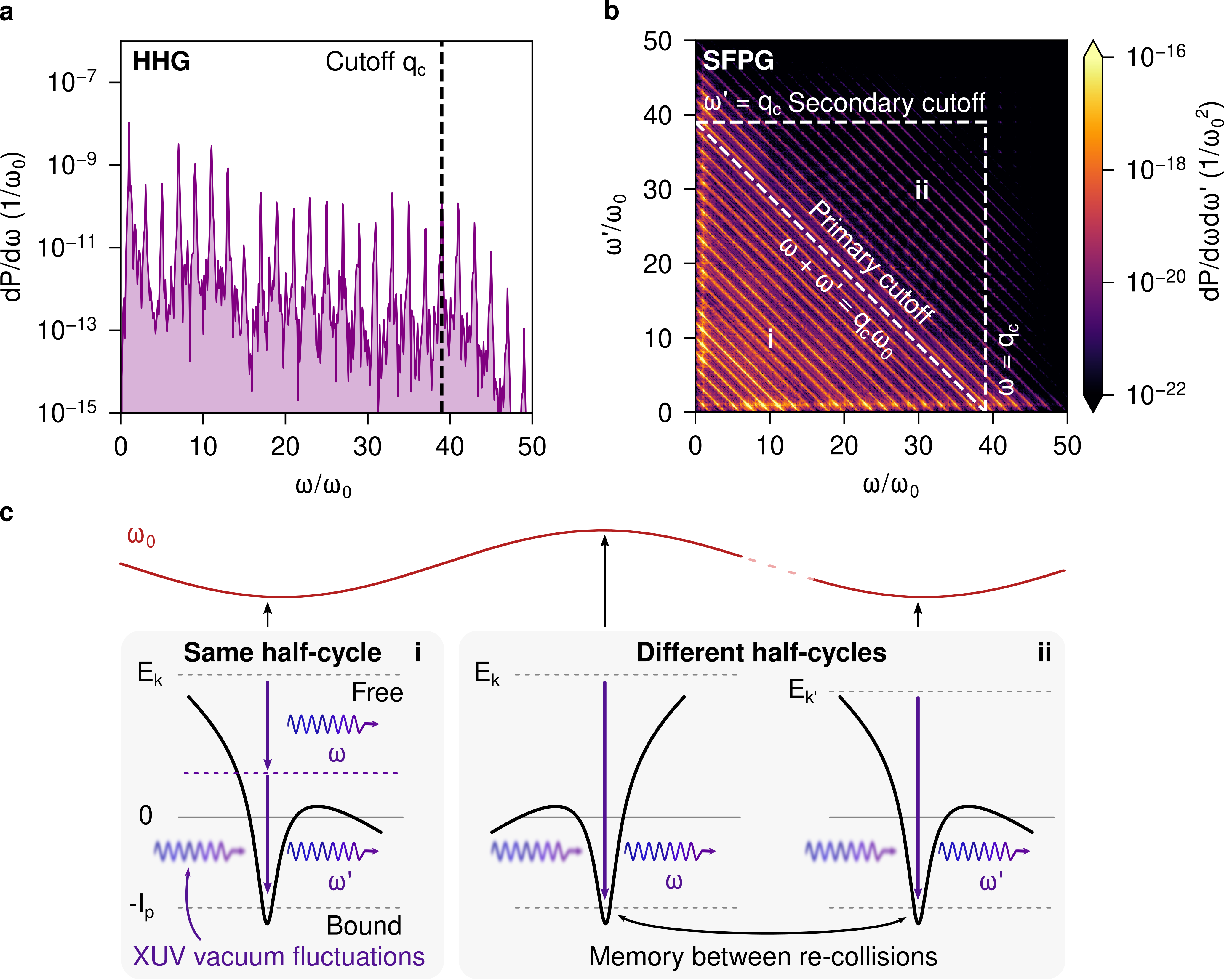}
    \caption{\textbf{Strong field pair generation from single atoms.} \textbf{(a)} HHG spectrum for a 1D model of Neon driven by $800$ nm radiation with intensity $I = 200$ TW/cm$^2$. The system exhibits a plateau over many harmonics, before reaching a cutoff at $q_c \approx 39$. \textbf{(b)} Differential emission probability of entangled pairs from a single particle for the system shown in (a). The frequency correlations satisfy $\omega + \omega' = q\omega_0$, where $q$ is an even integer. The pair emission spectrum exhibits two cutoff behaviors: (1) a primary cutoff corresponding to $\omega + \omega' = q_c\omega_0$, and (2) a secondary cutoff corresponding to $\omega, \omega' = q_c \omega_0$. \textbf{(c)} Three-step-model of entangled pair production from gases. Extreme ultraviolet vacuum fluctuations stimulate the production of entangled pairs at $\omega$ and $\omega'$. Panel (i) shows pairs emitted during the same re-collision. The maximum energy available for pair production is the maximum attainable value of $E_k + I_p$, leading to the primary cutoff in (b). Panel (ii) shows pairs emitted during different half cycles, leading to the secondary cutoff in (b). This correlated emission between different half cycles is indicative of a memory effect in the dipole.}
    \label{fig:single_particle}
\end{figure*}

\textbf{The concept of strong field pair generation:} The key ingredient for creating entangled pairs through SFPG is similar to that required for HHG: a sample illuminated by ultra-fast laser pulses with sufficient intensity to induce non-perturbative dynamics (Fig.~\ref{fig:schematic}a). From a quantum optical perspective (represented with Feynman diagrams in Fig.~\ref{fig:schematic}b), HHG is a non-perturbative process in which many photons at the drive frequency $\omega_0$ are converted into a high harmonic photon at frequency $q\omega_0$. 
In SFPG, the subject of this work, many driving photons at $\omega_0$ are converted into an entangled pair at frequencies $\omega$, $\omega'$, whose energies sum to some integer multiple $q$ of the drive ($\omega + \omega' = q\omega_0$). 

SFPG is in some ways analogous to spontaneous parametric down conversion (SPDC) in second-order nonlinear media (Fig.~\ref{fig:schematic}b). Both SPDC and SFPG yield entangled photon pairs which can exhibit correlations in frequency, angle, and polarization. However, as a strong field phenomenon, the highly non-perturbative nature of SFPG introduces significant differences from SPDC. For example, SPDC suffers from the constraint that to produce entangled photon pairs at a given frequency, one must pump with the sum of the frequencies to be emitted. This poses a significant challenge for entangling high frequency photons through SPDC, due to the scarcity of the intense high frequency sources, and intrinsic high frequency $\chi^{(2)}$ nonlinearities \cite{helk2021table} required to realize these effects. In contrast, the strong field nature of SFPG enables the emission of pairs at frequencies many times higher than that of the drive. In fact, the annihilation of dozens or hundreds of IR photons corresponds to emitted pair frequencies in the XUV and beyond (Fig.~\ref{fig:schematic}c). This is possible since the breakdown of perturbative nonlinear optics in strong-field settings enables high order $\chi^{(n)}$ processes to occur with an efficiency much higher than would be be possible otherwise.





\textbf{The physics of strong field pair generation:} We now explain the physical nature of SFPG, using a single 1D Neon atom driven at $\lambda_0 = 800$ nm as an example. Traditional HHG in such a system exhibits a characteristic ``plateau'' of harmonics generated with similar intensities, followed by a sharp ``cutoff,'' after which emission drops rapidly (Fig.~\ref{fig:single_particle}a). These behaviors are commonly described in terms of the so-called ``three-step-model,'', which describes the expedition of a valence electron in the driven atom in three critical steps: (1) tunnel ionization of the electron from its bound state into the continuum due to the strong field, (2) acceleration of the liberated electron in the strong field, and its eventual (3) re-collision with the parent ion, emitting high frequency radiation through recombination. 
The maximum possible energy gain results in a cutoff harmonic $q_c$ defined by $q_c \hbar \omega_0 \approx I_p + 3.17\,U_p$ \cite{corkum1993plasma, lewenstein1994theory}. Here, $I_p$ is the ionization energy, and $U_p = (e^2 E_0^2)/(2 m \omega_0^2)$ is the ponderomotive energy, where $e$ and $m$ are the electron charge and mass, and $E_0$ is the peak electric field.


HHG originates from the acceleration of the dipole moment $\braket{d(t)}$. For an atom driven by a linearly polarized field, the probability $P_{\text{HHG}}$ of HHG emission per unit frequency $\omega$ is
\begin{equation}
    \frac{dP_{\text{HHG}}}{d\omega} = \frac{2\alpha}{3\pi} \frac{\omega^3 |x(\omega)|^2}{c^2},
\end{equation}
where $\alpha$ is the fine structure constant, and $x(\omega)$ is the Fourier transform of the position matrix element $\braket{\psi_0|x(t)|\psi_0}$ taken on the initial electron state $\ket{\psi_0}$. This average dipole moment can be found through the Schr\"{o}dinger equation for the bound electron driven by a classical electromagnetic field. 

In contrast, SFPG originates from second-order dipole moment correlations $\braket{d(t)d(t')}$. We find that for a 1D model, the probability of SFPG emission $P_{\text{SFPG}}$ into frequencies $\omega$ and $\omega'$ is
\begin{equation}
     \frac{dP_{\text{SFPG}}}{d\omega d\omega'} = \frac{2\alpha^2}{9\pi^2} \frac{(\omega \omega')^3 |C_{xx}(\omega, \omega')|^2}{c^4},
\end{equation}
where $C_{xx}$ is the Fourier transform of a time-ordered ($\mathcal{T}$) connected correlation function $C_{xx}(t,t') = \braket{\psi_0|\mathcal{T} x(t) x(t') |\psi_0} - \braket{\psi_0|x(t)|\psi_0}\braket{\psi_0|x(t')|\psi_0}.$ The derivation of this result, as well as its generaliztion to 3D systems is provided in the S.I. Regardless of the dimensionality, the emitted spectrum of pairs exhibits a series of diagonal stripes, corresponding to the condition $\omega + \omega' = q\omega_0$ for integers $q$ (Fig.~\ref{fig:single_particle}b).



To connect SFPG to the three-step-model, we note that a number of past theoretical and experimental works have investigated HHG probed by weak XUV fields \cite{fleischer2008amplification, seres2010laser, kruger2019role}. These works revealed that even a very weak XUV probe impacts recombination (3) by weakly modulating the bound state as the free wavepacket collides with the parent ion \cite{fleischer2008generation}. 
Quantum optics provides the critical insight that even when no XUV probe field is applied, high frequency vacuum fluctuations persist; these XUV vacuum fluctuations virtually modulate the un-ionized portion of the bound electron, stimulating the production of pairs during electron recombination (Fig.~\ref{fig:single_particle}c, left). The primary contribution to SFPG is thus well-explained in terms of the three-step-model: The ionization (1) and acceleration (2) steps are unchanged. However, during vacuum-assisted recombination (3), the energy $E_k$ gained by the free electron is converted into an entangled pair ($\hbar(\omega + \omega') = E_k + I_p$). 

\emph{Cutoff behavior:} The physical picture provided above is closely tied to the question of whether or not SFPG also obeys a robust cutoff. Based on the relation of SFPG to the three-step-model described above, it is unsurprising that SFPG into frequencies $\omega$ and $\omega'$ exhibits a primary cutoff according to
\begin{equation}
    \omega + \omega' \approx q_c \omega_0,
    \label{eq:cutoff_1}
\end{equation}
where $q_c$ is the HHG cutoff harmonic. This is for the simple reason that the maximum energy which is ordinarily available for the production of a single HHG photon may now produce a pair. Frequencies within this primary cutoff lie in the lower triangular region of the pair spectrum (Fig.~\ref{fig:single_particle}b). 

\begin{figure*}
    \centering
    \includegraphics{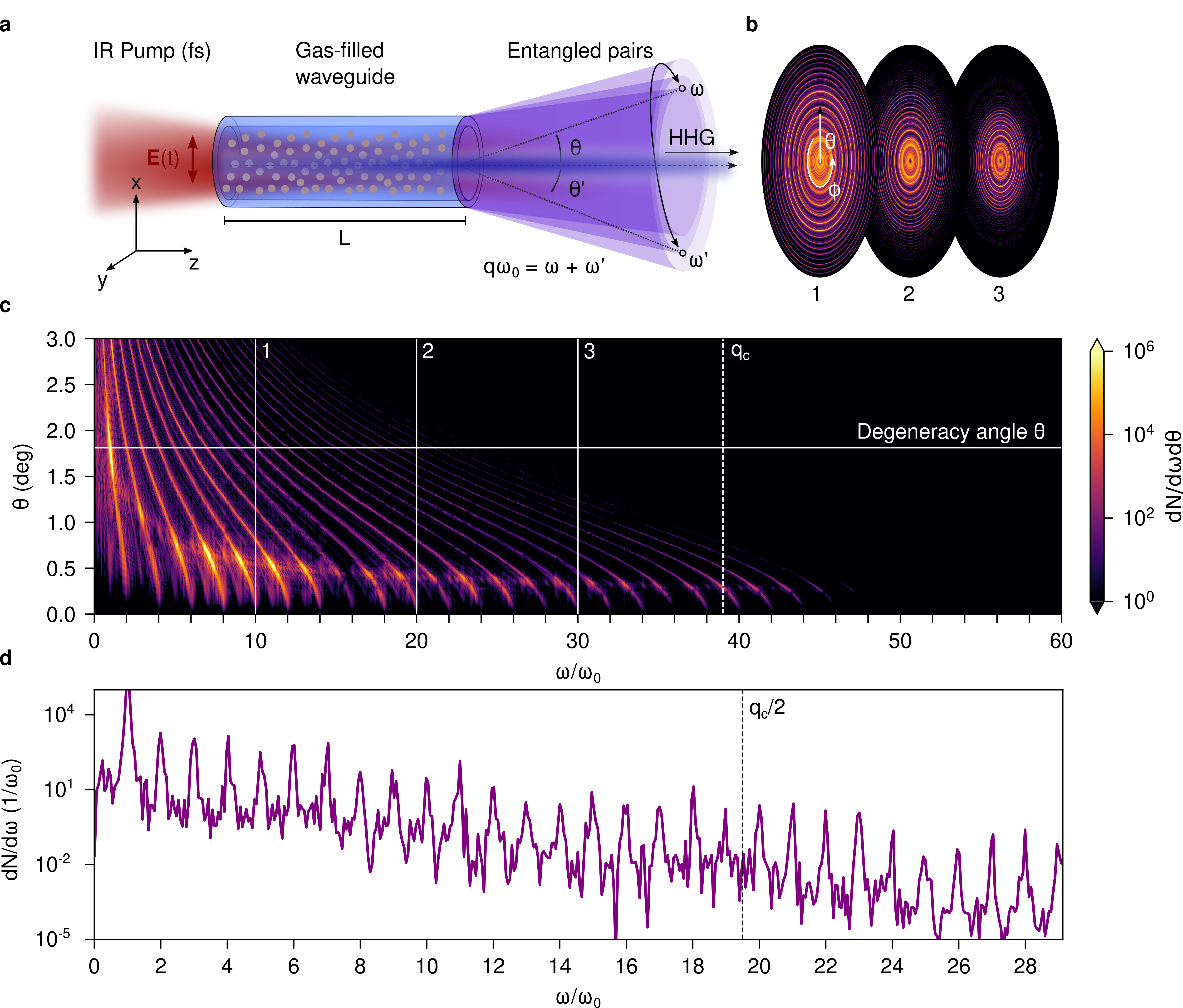}
    \caption{\textbf{Strong field pair generation from bulk gas samples.} \textbf{(a)} Gas filled waveguide pumped with femtosecond laser pulses. The ordinary HHG beam is produced on-axis along the $z$ axis. SFPG pairs of frequency $\omega$ and $\omega'$ are produced at angles $\theta$ and $\theta'$ off-axis. \textbf{(b)} Angular rings of radiation at fixed frequencies marked in (c). Each of the rings corresponds to a different integer order $q$. \textbf{(c)} Number of photons per shot $dN/d\omega d\theta$ produced per unit angle and frequency. Each of the stripes is the result of the angle-frequency correlations described by Eq.~\ref{eq:phase_match_simple} for even integers $q$. \textbf{(d)} Photon count rate per harmonic for measurement at the degeneracy angle $\theta_0$. Degenerate pairs are emitted most efficiently until the primary cutoff at $q_c/2$. Gas parameters are the same as those used in Fig.~\ref{fig:single_particle}. The gas sample is held in a hollow core waveguide which has length $L = 1$ mm, radius $a = 400 \mu$m, and is held at a pressure $P = 1$ atm.}
    \label{fig:bulk_emission}
\end{figure*}

However, the SFPG cutoff exhibits an additional complexity which is not possible in HHG. 
In particular, the two photon nature of SFPG presents the possibility that \emph{each photon of the pair is emitted during a different half-cycle of the drive} (Fig.~\ref{fig:single_particle}c, right). For such events, the electron gains energy during each of the distinct half-cycles so that each photon individually may be emitted up to the cutoff frequency. We refer to this behavior as the ``secondary cutoff,'' which is described by the constraint 
\begin{equation}
    \omega, \omega' \approx q_c \omega_0.
    \label{eq:cutoff_2}
\end{equation}
Contributions from these different half-cycle events result in a pair spectrum which does not drop off after the primary cutoff as rapidly as one might imagine. This presence of contributions beyond the primary cutoff (see upper triangle region of Fig.~\ref{fig:single_particle}b) indicates a notable memory effect in the temporal dipole correlations $C_{xx}(t,t')$. In other words, the recombination event associated with the first photon imprints a memory onto the bound state which can correlate with future recombinations. By performing a time-frequency analysis of the dipole correlations (see S.I.), we found that for this particular model, the strongest contribution beyond the primary cutoff stems from pair emissions correlated between neighboring half-cycles.

\emph{Selection rules:} A famous feature of traditional HHG is that an inversion-symmetric sample driven by a linearly polarized laser field of a single frequency emits only odd harmonics. Various symmetry-breaking techniques \cite{liu2017high, kim2005highly, neufeld2019floquet, tzur2022selection}, 
have been employed to control the selection rules for frequencies and polarizations. We found that similar constraints govern SFPG. For the simple case of a symmetric potential driven by a monochromatic, linearly polarized field, SFPG pairs are subject to the constraint $\omega + \omega' = q\omega_0$, where $q$ is an even integer, rather than an odd integer. This is consistent with the famous result of perturbative nonlinear optics that centro-symmetric materials have no even-order nonlinearities ($\chi^{(2)}$ for example) \cite{boyd2020nonlinear}. In such a symmetric material, SPDC ($q=1$) is forbidden, but spontaneous four-wave-mixing ($q=2$) is allowed. It follows that breaking dynamical symmetry would enable SFPG processes with all integers $q$. We anticipate that further analysis of spatio-temporal symmetries will lead to robust SFPG selection rules.

\emph{Efficiency of SFPG:} A key area of interest is the efficiency of SFPG compared to ordinary HHG. At the single particle level, SFPG process is strictly less efficient than HHG, since HHG occurs at first order in the emitted field, while SFPG occurs at second order (Fig.~\ref{fig:schematic}b). For a given single particle sample and driving field, it is loosely the case that the probability of a pair emission event (SFPG) at frequencies $\omega$ and $\omega'$ is on the order of the product of the probabilities for single photon emission events (HHG) at the two frequencies. In noble gasses, HHG probabilities for a single harmonic typically range from $10^{-10}$ -- $10^{-6}$. Correspondingly, pair generation probabilities can take values in the approximate range of $10^{-20}$ -- $10^{-12}$. In the next section, we show that even though the per-particle emission probability is low, phase matched interactions can lead to detectable numbers of pairs which are distinguishable from HHG. 

\begin{figure*}
    \centering
    \includegraphics{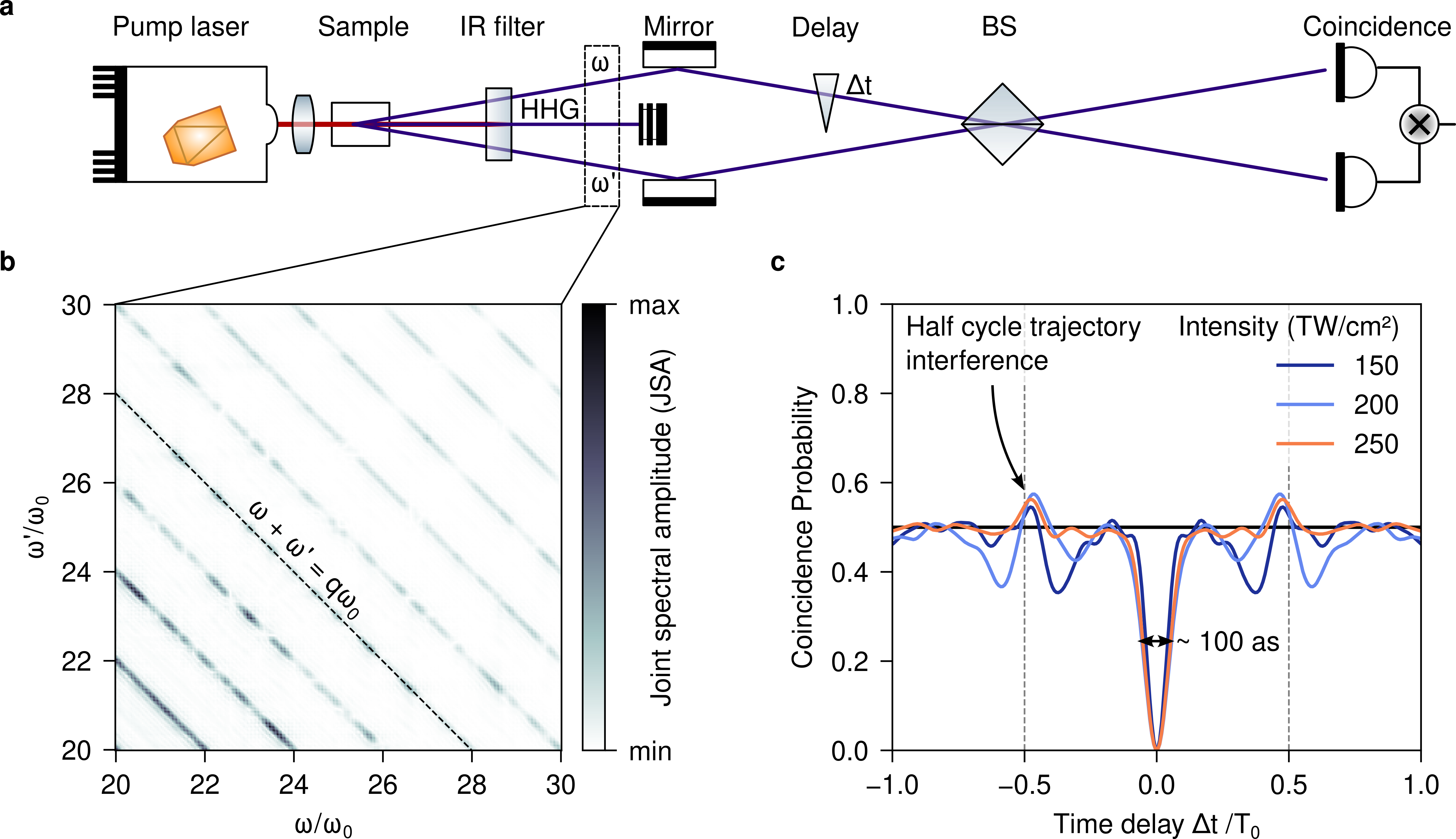}
    \caption{\textbf{Quantum nature of strong field pair generation.} \textbf{(a)} Schematic of a Hong-Ou-Mandel (HOM) experiment for SFPG pairs. Pairs are collected from two arms, and then directed through a beam-splitter (BS), and to a coincidence detection apparatus. One of the arms experiences a time delay $\Delta t$. \textbf{(b)} Joint spectral amplitude which shows the quantum state of the collected pairs. Many stripes of frequency anti-correlation are present with comparable amplitude. \textbf{(c)} Attosecond HOM dip at different driving intensities, obtained from the coincidence probability from the setup depicted in (a). Varying the driving intensity impacts the fringes which result from interference between different integer orders $q$.}
    \label{fig:quantum_state}
\end{figure*}

\textbf{Correlations in angle and frequency:} Let us detail how SFPG can be realized on mature HHG platforms. For concreteness, we focus on a hollow-core waveguide filled with a noble gas with variable pressure used for phase matching control (Fig.~\ref{fig:bulk_emission}a). Our theory allows for the computation of the numbers of pairs per solid angle and frequency which are emitted from a uniformly illuminated gas sample, taking into account the dispersion of the pump and signal frequencies (see S.I.). We found that for a 1D model, SFPG signal consists of entangled pairs directed into cones at different angles, carrying some features similar to SPDC. The classically measured angle and frequency spectrum of SFPG pairs exhibits many stripes of correlation between angle and frequency, which arise from energy and momentum conservation constraints (Fig.~\ref{fig:bulk_emission}c). In particular, if the emitted photons at $\omega$ and $\omega'$ have the same refractive index $n$, then the emission angle satisfies
\begin{equation}
    \cos\theta = \frac{2 n^2 \omega - n^2 q \omega_0 + n_0^2 q \omega_0}{2 n n_0 \omega},
    \label{eq:phase_match_simple}
\end{equation}
where $n_0$ is the index at the pump frequency $\omega_0$. This constraint between angles and frequencies is the origin of the stripes seen in Fig.~\ref{fig:bulk_emission}c. 

For the parameters considered, the pairs are emitted a few degrees away from the pump axis. At a fixed frequency of observation, emission is peaked at many discrete angles, corresponding to different integer orders $q$ (Fig.~\ref{fig:bulk_emission}b). Similarly, at a fixed angle of observation, emission is peaked in a frequency comb pattern (Fig.~\ref{fig:bulk_emission}d). Emission can be either degenerate ($\omega = \omega'$), or highly nondegenerate $(\omega \neq \omega')$, and both contributions are substantial. Additionally, the spectrum of SFPG collected over many angles or frequencies will be broad, since it is the correlation spectrum, not the emitted frequencies themselves, which exhibit harmonics.


An important practical aspect of many nonlinear optical processes is phase-matching, which equivalently corresponds to energy and momentum conservation of incoming and outgoing fields. For ordinary HHG, phase matching amounts to index matching the driving and emitted frequencies (i.e., $n(\omega_0) = n(\omega)$). For SFPG, the considerations are different: the phase velocity of the pump needs to exceed that of the emitted frequencies, giving a Cherenkov-type condition ($\cos\theta_0 = n(\omega_0)/n(\omega)$). The higher the pump phase velocity compared to that of the emitted frequencies, the larger the angle the phase-matched pairs will make with the optical axis. For these reasons, \emph{SFPG phase matching will be ideal for parameters where ordinary HHG is totally non-phase-matched.} A detailed analysis of phase matching reveals that SFPG can be perfectly phase matched while HHG phase mismatch provides orders of magnitude of suppression (see S.I.). For these reasons, it may actually be desirable to operate with higher gas pressures and higher ionization fractions to increase the pump phase velocity. The ability to use higher gas pressures $P$ is also favorable from an efficiency point of view due to the $\propto P^2$ scaling of yield which comes from the density of atoms.

\textbf{Quantum states of light from SFPG:} The quantum state of light produced by SFPG features a rich structure of entanglement, primarily due to the possibility of pair emission from many integers $q$ simultaneously. This is most easily seen through the joint spectral amplitude for detecting a pair of photons at $\omega$ and $\omega'$, which exhibits many stripes of anti-correlation (Fig.~\ref{fig:quantum_state}b). In contrast, the joint spectral amplitude produced from SPDC consists only of a single stripe of anti-correlation ($q=1$). We now highlight three essential features of the quantum state of SFPG pairs.

First, the attosecond pulse nature of the entangled pairs is made evident through a hypothetical Hong-Ou-Mandel (HOM) experiment (Fig.~\ref{fig:quantum_state}a) that counts coincident detection of photons as a function of time delay between arms \cite{hong1987measurement}. The resulting HOM curves for various intensities all show a characteristic dip at zero time delay (Fig.~\ref{fig:quantum_state}c). These dips occur over a $\sim$100 as timescale, owing to the large frequency bandwidth covered by the phase matched pairs. 
In addition to the zero delay HOM dip, all three driving intensities produce a more complex structure of interference fringes, which share a common strong feature at a half cycle delay. We attribute this feature to the memory effect between neighboring half cycles (Fig.~\ref{fig:single_particle}c, panel (ii))

Second, the SFPG state can be used to create a heralded single photon attosecond pulse. The fact that many different numbers of pump photons can be converted into an entangled pair means that a photon measured at frequency $\omega$ has many possible partner photons at $\omega' = q\omega_0 - \omega$. Thus, by measuring one photon, the other heralded photon lies in a coherent superposition of many frequencies spaced by $\omega_0$. This can be most easily seen by taking a ``slice'' of the joint spectral amplitude at some fixed frequency. Since this heralded photon consists of many frequencies in the XUV spaced by $\omega_0$, the heralded photon is an \emph{attosecond pulse train carried by a single photon.} These high frequency heralding experiments could be conducted using similar techniques to those used to herald hard X-rays from facility-scale X-ray PDC sources \cite{strizhevsky2021efficient}. 

Finally, the states generated by SFPG also present considerable interest from the standpoint of entanglement structure and quantum information. The simultaneous presence of many integers $q$ in the joint spectral amplitude gives a richer structure than that seen in traditional SPDC. We quantified this structure through calculations of Schmidt number and entanglement entropy for the entangled states, indeed indicating a high degree of entanglement (see S.I.). 
A SFPG source based on strongly driven gases also has the unique feature that by tuning the gas pressure, one affects the phase matching, which in turn controls the quantum state and corresponding entanglement. Accordingly, we foresee that these high dimensional entangled states may serve as a platform to represent high dimensional quantum information in the high frequency and attosecond regimes. Although these high frequencies carry the disadvantage that constructing optical elements can be difficult, they come with the advantage of very high efficiency detection --- a critical element for quantum state characterization.

\section{Factors for experimental observation}

We now outline some key factors for experiments on SFPG. For the parameters we considered (Fig.~\ref{fig:bulk_emission}), many thousands of entangled pairs are created in a single shot, with hundreds or more over a 1 eV bandwidth in some cases. If the incoming pulse which contains $N_{\text{in}} \approx 10^{17}$ IR photons creates $N_{\text{out}} \approx 100$ photons over some narrow XUV bandwidth, this corresponds to an efficiency of $N_{\text{out}}/N_{\text{in}} \approx 10^{-15}$. We thus conclude that SFPG should be efficient enough to enable observation, even if counts are at the single photon level \cite{fuchs2022photon}. 

To aid observation, it will be crucial to create frequency and/or angular ranges where SFPG counts exceed any HHG background. This can be done in part by exploiting the differences in angular distribution between HHG and SFPG. In general, momentum conservation dictates that ordinary HHG is emitted along the pump axis, with narrow angular divergence of only a few milliradians in some cases \cite{abbing2020divergence, rego2022necklace}. Conversely, SFPG pairs have the potential to be emitted in cones which are tens to hundreds of milliradians off axis, spatially distinguishing them from HHG. 


Another helpful factor is that over a narrow angular range, SFPG can produce pairs peaked at frequencies which are not permitted from ordinary HHG (see Fig.~\ref{fig:bulk_emission}c,d). In particular, measurement at the appropriate angle $\theta_0$ for degenerate pairs will result in the detection of both even and odd harmonics in a spatially symmetric noble gas, in contrast to the strictly odd pairs which are allowed for ordinary HHG. More strikingly, measurement of non-degenerate pairs could lead to the observation of frequencies which are not harmonics at all. 

Yet another important consideration is that ordinary HHG will be entirely mismatched when SFPG is optimally matched. This factor could provide 4-5 orders of magnitude of suppression of the background HHG signal compared to its ordinary optimal strength. It may also be possible to engineer more complex HHG geometries (such as two pump beams which come in at different angles) to force the SFPG beams to emerge at entirely different angles from the HHG signal. Once a SFPG signal is isolated, it may be possible to use an interferometry scheme such as the one recently reported in \cite{harrison2023generation} to discern the quantum nature of the pairs.

Finally, while we have focused in this work on SFPG from gas samples, much of the fundamental physics in SFPG should carry over into solids used for HHG, frequently driven with mid-IR pulses \cite{ghimire2019high}. In principle, solids could be used to realize non-perturbative conversion of many mid-IR photons into pairs ranging from the optical to UV. These platforms offer the additional advantage of the potential to engineer metasurfaces or other optical structures which could offer degrees of angle and frequency control over emitted pairs \cite{santiago2022resonant}.


\section{Conclusion and outlook}

We have presented the theory of strong field pair generation (SFPG): a non-perturbative nonlinear optical process in which many photons of a high intensity driving field incident on matter are converted into an entangled pair of photons at high frequencies. Such sources have a potential to generate both degenerate and highly non-degenerate entangled photon pairs, covering large swaths of the electromagnetic spectrum over which entangled pairs have never been produced. This method also carries the distinct advantage that it can generate these pairs using mature HHG platforms, and without reliance on a pre-existing source of intense high frequency radiation.

The generation of entangled pairs at high frequencies can enable critical new applications in XUV/X-ray quantum optics. As one example, sources in the water window could produce new modalities of quantum microscopy which exploit correlated pairs to image biological samples with improved phase sensitivity. As another example, the generation of highly non-degenerate pairs consisting of XUV radiation entangled with infrared radiation could yield a quantum optical interface between optical and XUV radiation. Moreover, this work paves the way toward further studies of dipole correlations in strongly time-driven systems. In the future, the measurement of these correlated pairs may provide a lens into aspects of the attosecond dynamics of matter, such as correlation and memory effects, which are not accessible through classical detection schemes. 

The eventual experimental realization of SFPG pairs is expected to serve as an important milestone for the broader challenge of bringing quantum optics to the XUV and soft X-ray regimes. We anticipate that such an observation would open new domains of research in attosecond science, and provide important fundamental tests for the nascent field of strong-field quantum optics.

\section{Acknowledgments}

J.S. acknowledges prior support of a Mathworks graduate fellowship, as well as prior support of a National Defense Science and Engineering (NDSEG) fellowship.  This research was supported by Grant No 2022144 from the United States-Israel Binational Science Foundation (BSF). This research project was partially supported by the Helen Diller Quantum Center at the Technion through the Flagship research project (QUBIT).This material is based upon work sponsored in part by the U.S. Army DEVCOM ARL Army Research Office through the MIT Institute for Soldier Nanotechnologies under Cooperative Agreement number W911NF-23-2-0121, and also supported in part by the Air Force Office of Scientific Research under the award number FA9550-21-1-0299.

\bibliographystyle{unsrt}
\bibliography{bibliography.bib}

\end{document}